\documentclass[conference]{IEEEtran}
\IEEEoverridecommandlockouts
\usepackage{cite}
\usepackage{amsmath,amssymb,amsfonts}
\usepackage{algorithmic}
\usepackage{graphicx}
\usepackage{textcomp}
\usepackage{xcolor}

\newcommand{\fref}[1]{Fig.~\ref{#1}}

\def\BibTeX{{\rm B\kern-.05em{\sc i\kern-.025em b}\kern-.08em
    T\kern-.1667em\lower.7ex\hbox{E}\kern-.125emX}}
\begin{document}

\title{A Bidirectional Power Router for\\Traceable Multi-energy Management
\thanks{This work was partially supported by JSPS KAKENHI Grant No.23K26128, No.24K17262, and No.24K07496.}
}

\author{\IEEEauthorblockN{Shiu Mochiyama}
\IEEEauthorblockA{\textit{Dept. of Electrical Engineering} \\
\textit{Kyoto University}\\
Kyoto, Japan \\
ORCID: 0000-0001-7126-9187}
\and
\IEEEauthorblockN{Ryo Takahashi}
\IEEEauthorblockA{\textit{Dept. of Mechanical and Electrical Systems Engineering} \\
\textit{Kyoto University of Advanced Science}\\
Kyoto, Japan \\
ORCID: 0000-0001-8592-3856}
\and
\IEEEauthorblockN{Yoshihiko Susuki}
\IEEEauthorblockA{\textit{Dept. of Electrical Engineering} \\
\textit{Kyoto University}\\
Kyoto, Japan \\
ORCID: 0000-0003-4701-1199}
}

\maketitle

\begin{abstract}
To address challenges in improving self-consumption of renewables and resilience in local residential power systems, the earlier work of the authors introduced a novel multi-energy management concept, integrating bidirectional power routing and electricity-hydrogen conversion.
This paper focuses on an experimental verification of the bidirectional power router based on line-switching, the essential hardware to realize the concept. 
The primary contribution is the validation of the router's capability to handle dynamic change of bidirectional power flow. 
Furthermore, to achieve bidirectional power routing without affecting the smooth and stable operation of the power system, a novel algorithm for router's switching is designed based on power flow monitoring. 
The effectiveness of the proposed method is demonstrated through an experiment using a setup with a commercially available stationary battery.
\end{abstract}

\begin{IEEEkeywords}
bidirectional power routing, multi-energy system, self-consumption, resilience
\end{IEEEkeywords}

\section{Introduction}
\label{sec:intro}

\subsection{Background}

This study focuses on an energy management system for a local residential community. 
Such an energy system faces the following urgent challenges in terms of carbon neutrality and resilience. 

Although there is no doubt that the mass introduction of photovoltaic (PV) generation to households is a key to reducing carbon emissions, its full utilization is not straightforward. 
The uncertain and fluctuating output of PV does not always match the consumption profile of households, and flowing excess energy back to the grid is sometimes prohibited to avoid disturbing the grid's stable operation\cite{Liang-2017}. 
These circumstances have led to a demand for technologies that can increase the rate of self-consumption.

Another challenge is to improve the resilience of the power system. 
In addition to reducing grid dependency by improving self-consumption during normal operation, it is necessary to prepare for more severe conditions, such as natural disasters, where grid power is largely lost. 
In such cases, consumers should be supplied according to their priority.
Unfortunately, the current power system does not cover such operation%
\footnote{As a recent example, in the event of severe weather forecast, the Japanese government issued multiple requests of a \textit{voluntary} action to save use of electricity \cite{TheJapanTimes-2023}.}.

From the perspective of quantitative evaluation of carbon neutrality, making the origin of energy traceable is an important technical challenge. 
Similarly to the certification of green hydrogen\cite{Gregor.Svensson-2023}, establishing a technology to physically distinguish electricity according to its primary energy source will enable an accurate evaluation of the environmental impact. 
In particular, in multi-energy systems\cite{Geidl.Andersson-2007,OMalley.Kroposki-2013,Wu.etal-2016,Mancarella-2014,Chertkov.Andersson-2020} that include the mutual conversion of hydrogen and electricity, it is necessary to be able to trace the origin of all secondary energy sources.

\subsection{Energy Management with Bidirectional Power Routing}

\begin{figure}
    \centering
    \includegraphics[width=0.8\linewidth]{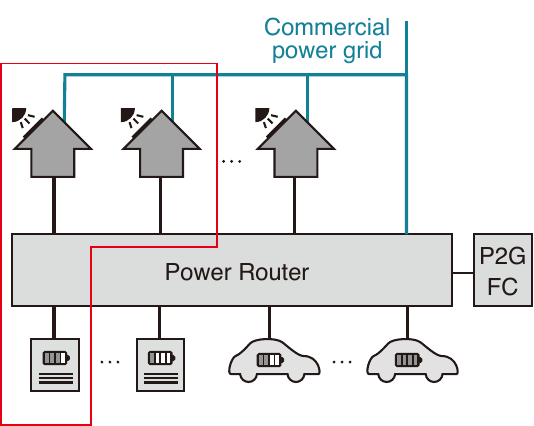}
    \caption{Overview of the energy management system considered in this study. }
    \label{fig:concept}
\end{figure}

To address the above challenges, the authors' group has been working on a novel multi-energy management system integrating bidirectional electric power routing and hydrogen-electricity conversion\cite{Mochiyama.etal-2024}. 
\fref{fig:concept} depicts an overview of the energy management system. 
The target is a residential community consisting of houses, batteries, a power router, and a shared facility for hydrogen-electricity mutual conversion based on power-to-gas (P2G) and fuel cell (FC) technologies\cite{Staffell.etal-2019}. 
Each house is assumed to be equipped with a PV generation. 
The batteries can be either installed on electric vehicles or stationary ones. 
The hydrogen facility serves as a larger and longer-term energy storage than batteries\cite{Evans.etal-2012}. 
Then, the houses, batteries, and hydrogen facility are connected to the power router. 
The role of the power router is to create their physical connection as an electrical circuit. 
The router hardware consists of a crossbar switch circuit that connects two ports to form a closed circuit between them and to isolate them from others\cite{Takuno.etal-2010}. 

The introduction of power routing enables physically distinguishable and traceable transactions of electricity between a battery and a house or the hydrogen facility. 
In the following, we summarize on the basis of the discussion in the earlier publication\cite{Mochiyama.etal-2024} how these features address the aforementioned challenges.

First, the self-consumption of renewables is increased by creating a sharing economy of batteries\cite{Kalathil.etal-2019}. 
By sharing renewable generation and storage capacity among many households, it is possible to improve self-consumption as a whole community without installing an extremely large battery in each household.
Here, the traceability of power is the key to putting the concept into practice. 
Physically distinguishing power flows by power routing enables us to tell precisely which portion of the stored energy belongs to whom. 

Second, integrating the hydrogen facility with power routing enhances resilience in emergencies. 
When grid power is significantly or entirely limited, the hydrogen facility acts as a primary energy source by utilizing stored hydrogen. 
Then, the selective power flow management via the power router ensures prioritized distribution of power to crucial loads.

Third, power routing facilitates the traceability of the electricity's origin to assess carbon neutrality correctly. 
Especially, the selectivity of devices connected to the P2G allows us to identify the origin of electricity used to produce hydrogen. 

\subsection{Contributions}

The energy management system comprises both a software technology, which determines how to coordinate the multiple batteries to obtain the best result, and a hardware technology, which puts into practice the optimized transactions of electricity. 
This study focuses on the latter, providing experimental verification of the essential function of the bidirectional power router hardware.
The main contributions of the paper are summarized as follows. 
\begin{itemize}
    \item We verify the bidirectional routing in a dynamic condition. 
    Here, ``dynamic" refers to the situation where the direction of power flow through a router's port changes in a single continuous experiment. 
    In the previous report\cite{Mochiyama.etal-2024}, bidirectional routing was verified by conducting separate experiments for positive and negative power flows with all devices temporarily stopped during the switchover. 
    Since battery operation in general includes dynamic changes in the direction of power flow, the results of this paper demonstrate the feasibility of the proposed system concept in a realistic condition. 

    \item As a core technique to realize the bidirectional power routing in a dynamic condition, we propose a method for circuit switching of the router that minimizes the undesired effect on the bus voltage and the operation of devices connected to the router. 
    The proposed method switches the connection relationship considering the transient behavior of the connected devices, e.g. rate-limited change of the output power of a battery. 
    We utilize the power sensing function of the router introduced in the previous study\cite{Mochiyama.etal-2024} to monitor the transient state of connected devices and perform the switching at the appropriate timing. 
    The proposed method is verified through experiments using a real device, a stationary battery that is in practical use in Japan. 
\end{itemize}

\section{Energy Management System with Bidirectional Power Router}

\subsection{Principle Operation of Energy Management System}

In the energy management system, controllable elements are the charging or discharging power of the batteries and the physical connection of the router ports. 
Here, it should be noted that individual batteries may be subject to restrictions on charging and discharging due to reaching their maximum capacity or being absent for travel (in the case of onboard batteries), for example. 
Still, it is reasonable to assume that their availability can at least roughly be predicted in advance. 
The household consumption and PV output are also predictable to some extent but not precisely, and are not controllable in general. 
After all, the operation of the system involves setting references in advance for both the router ports' connection and output power of the batteries by optimization based on various forecasts etc., and controlling the actual output of the batteries in real time taking into account fluctuations in PV power generation and household demand.

In the remainder of this paper, we focus on the verification of the router capability to achieve the above operation. 
We omit the optimization process, which will be discussed in a future publication. 
In the verification of this paper, we assume that the references are already somehow obtained. 

\subsection{Hardware of Router}
\label{ssec:router}

The router realizes the physical connection (or disconnection) of the ports. 
The essential parts of the router are a crossbar-type switching matrix and a sensor unit for monitoring the power flow. 
In this paper, we use the same hardware as developed in the previous study\cite{Mochiyama.etal-2024}. 
In the following, we briefly review the hardware configuration of the router. 

\begin{figure}
    \centering
    \includegraphics[width=0.85\linewidth]{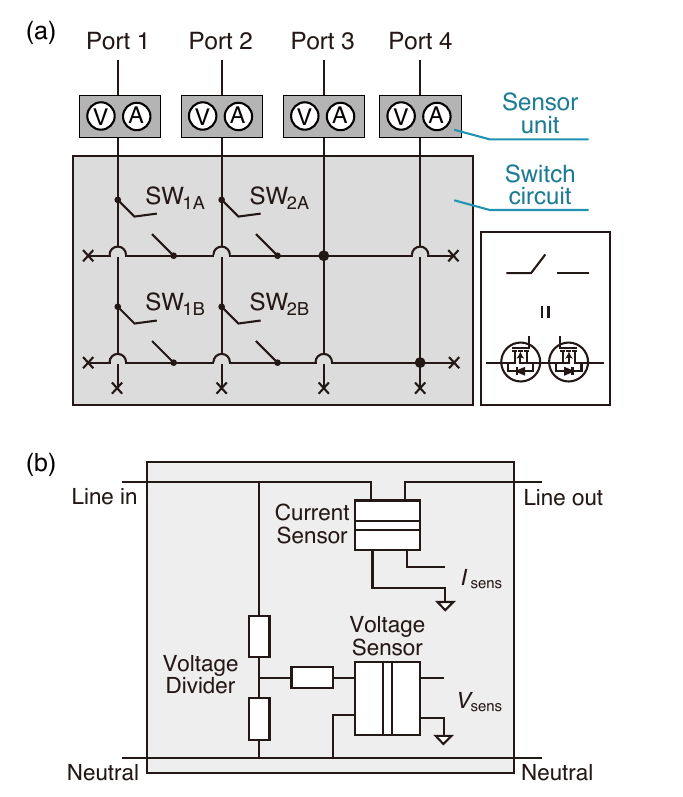}
    \caption{Configuration of the hardware of the power router developed in the previous work\cite{Mochiyama.etal-2024}. (a) Overall circuit. (b) Detail of the sensor unit. Reprinted from \cite{Mochiyama.etal-2024} with modifications. }
    \label{fig:router}
\end{figure}

\fref{fig:router} depicts the configuration of the router. 
The switching circuit comprises four bidirectional switches, each of which is a back-to-back configuration of two SiC MOSFETs of $650\,\mathrm{V}$, $20\,\mathrm{A}$ ratings (C3M0060065; Wolfspeed, Inc.). 
For example, turning on the switch $\mathrm{SW_{1A}}$ connects port 1 and port 3. 
The gate signals for the switches are made at the router's controller. 
The switching matrix is a subset of the full configuration that comprises eight switches and is sufficient for the purpose of this paper. 

Each port is equipped with a sensor unit that can measure instantaneous values of voltage and current. 
The unit comprises a voltage divider with an isolated voltage sensor IC and a hall effect current sensor IC. 
Note that \fref{fig:router}~(b) omits some auxiliary components used, for example, for signal filtering and IC powering to enhance the visibility.
The obtained values are transferred to the router's controller and processed in a real-time manner to monitor the power flow at each port. 

Here, it should be emphasized that the key features discussed in Section~\ref{sec:intro} are only achieved altogether with the power router hardware based on line-switching. 
There are also proposals of power routers based on multi-stage power conversion\cite{Huang.etal-2011,Abe.etal-2011,Stalling.etal-2012,Kado.etal-2016a,Yu.etal-2023,Liu.etal-2024a}
(see \cite{Mochiyama.etal-2024} for a detailed comparative review).
They basically comprise multiple power conversion elements per port, although there are variations in their detailed implementation such as the type of their internal bus (AC or DC). 
Power routers based on power conversion can manage variable voltage levels and both AC and DC inputs, allowing for connections to kV-level distribution lines, consumer loads, and DC supplies from renewables. 
In fact, a literature\cite{Yu.etal-2023} discusses the application of this function to a multi-energy system of hydrogen and electricity. 
Nonetheless, the power routers based on power conversion do not completely meet our aim, physically distinguishing power flow with traceability. 
For this reason we focus on the router based on line-switching. 

\subsection{Switching Algorithm}

One of the contributions of this paper is a proposal of an algorithm for the router switching based on the power flow monitoring. 
The proposed algorithm controls the timing of circuit switching based on the instantaneous power output of the connected devices. 
As mentioned above, we assume that the commands for the router's switching state and the output power reference for the batteries are given in advance. 
These commands are set for each unit time interval of optimization, so they undergo a step change between the intervals.
However, the battery output does not respond immediately to the command values, but has delay and lag. 
If the router switches immediately at the beginning of an interval without considering this response, unexpected behavior may occur in the battery and the devices connected to the house. 
In particular, when the output polarity (i.e., charging or discharging) of the battery differs before and after the switch, proper operation is not expected. 
The proposed algorithm provides a method to determine an appropriate switching timing, taking into account the response of the battery output. 

\begin{figure}
    \centering
    \includegraphics[width=0.9\linewidth]{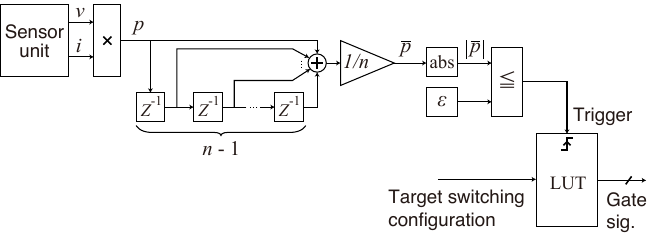}
    \caption{Simplified block diagram of the proposed switching algorithm. }
    \label{fig:alg}
\end{figure}

\fref{fig:alg} illustrates the overview of the proposed algorithm. 
The main point is to switch the circuit after confirming that the output power of the connected device has reached (almost) zero, rather than immediately after the mode switch command is issued. 

Zero power detection is implemented by comparing the absolute value of averaged power with $\varepsilon$, a sufficiently small number%
\footnote{We use $\varepsilon$ instead of exact zero for comparison because the measurement data is available at discrete sample times. For example, if the comparison is performed with exact zero, the zero cross of the averaged power cannot be detected when the sign changes between two consecutive samples.}. 
Simple moving average of $n$ samples, where $n$ is chosen so that the averaging window becomes an integer multiple of the period of AC bus voltage. 

The detection of (almost) zero power triggers the change of the gate signals given to the router. 
The gate signals are determined by a predefined lookup table, where the target of router's switching configuration is associated with the on/off states of the switches. 

\section{Experimental Verification}

\subsection{Setups}

We set up an experimental system corresponding to the area enclosed by the red line in \fref{fig:concept}. 
This three-port configuration is necessary and sufficient for verifying the capability of the bidirectional router hardware. 
This is because, based on the operating principle of the router, each pair of a battery and a house forms an independent circuit from others. 
If we verify dynamic switching with a focus on a particular battery, the same applies for other pairs. 

\begin{figure}
    \centering
    \includegraphics[width=0.75\linewidth]{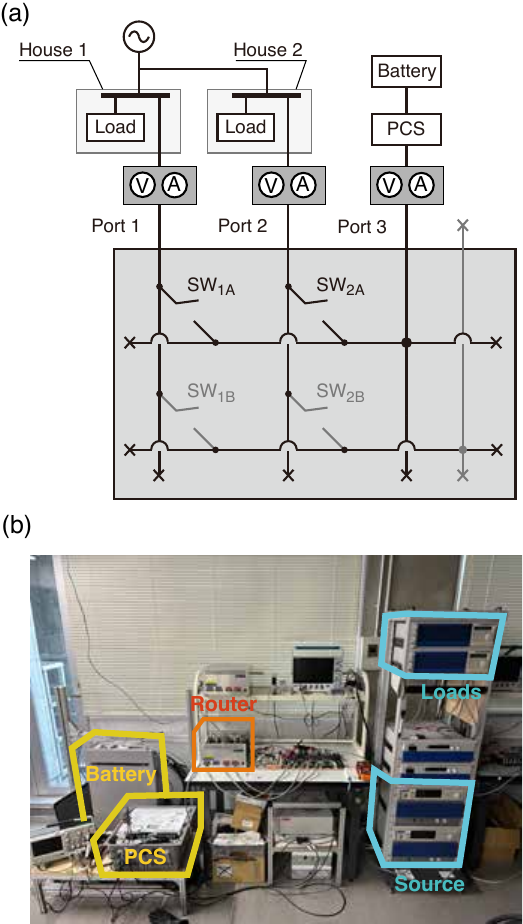}
    \caption{Experimental setup for the verification of bidirectional power routing. (a) Single-line diagram of the setup. (b) Photograph of the setup. }
    \label{fig:exp-setups}
\end{figure}

\fref{fig:exp-setups} depicts the configuration of the experimental setup. 
Port 1, 2, and 3 of the router are connected to house 1, house 2, and battery, respectively. 
Each house is modeled by a regulated power supply (PCR1000LE; Kikusui Electronics Corp.) and an electronic load (PCZ1000A; Kikusui Electronics Corp.). 
One power source is connected common to the two houses since it models the commercial grid. 
The source is set to output a single-phase $200\,\mathrm{Vrms}$ ac.
The electronic loads represent the sum of consumption in the individual households. 
The battery and PCS originate from a commercially available stationary energy storage system designed for general household use in Japan. 
The battery (KP-BU65B-S; Omron Social Solutions Co., Ltd.) features a capacity of $6.5\,\mathrm{kWh}$. 
Combined with the PCS (KPBP-A; Omron Social Solutions Co., Ltd.), they provide a single-phase output of $200\,\mathrm{Vrms}$ with synchronization to the grid-emurating source. 

The router controls the connection between battery and one of the houses. 
Since we use three of the four ports, the necessary connection relationship can be achieved by controlling the two switches $\mathrm{SW_{1A}}$ and $\mathrm{SW_{2A}}$. 
Other switches are left open. 

\subsection{Scenario}

We consider the transition between the following two operation conditions of the router and the battery. 

In the time before the transition, we assume that house 1 has less demand than PV's output and house 2 has a balanced supply and demand. 
The target of power management is supplying $700\,\mathrm{W}$ from house 1 to the battery. 
Then, the corresponding battery command is $700\,\mathrm{W}$ charging, and the switch states of the router are: $\mathrm{(SW_{1A},SW_{2A})=(ON,OFF)}$. 
We denote this router state by mode 1. 

In the time after the transition, we assume that house 1 reaches a balanced supply and demand and house 2 turns to have more demand than PV's output. 
The target of power management is supplying $700\,\mathrm{W}$ from battery to house 2. 
The corresponding battery command is $700\,\mathrm{W}$ discharging, and the switch states of the router are: $\mathrm{(SW_{1A},SW_{2A})=(OFF,ON)}$. 
We denote this router state by mode 2. 

\subsection{Results and Discussions}

\begin{figure}
    \centering
    \includegraphics[width=.9\linewidth]{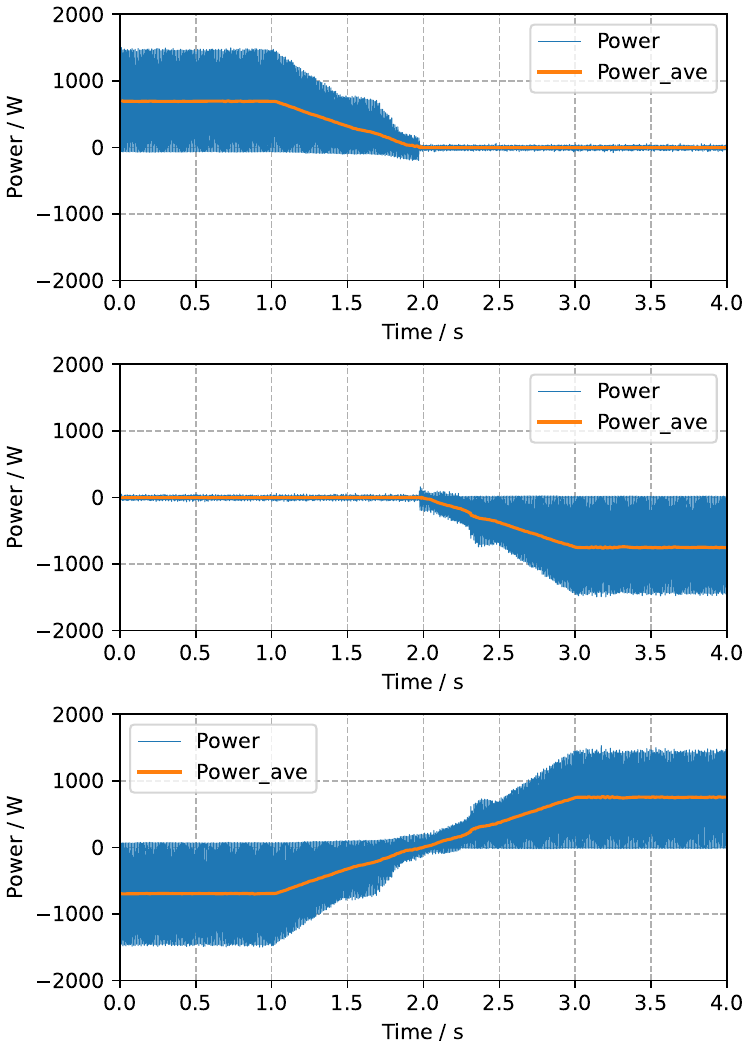}
    \caption{Instantaneous and averaged power waveforms measured at the router's three ports. Top, middle, and bottom depict the measurements in port 1, 2, and 3, respectively. In each, the blue line represents the instantaneous power calculated by the product of the voltage and current measurements, and the red line represents the averaged power calculated by taking a moving average of the instantaneous power. }
    \label{fig:p123}
\end{figure}

\fref{fig:p123} presents the instantaneous and averaged power waveforms measured at the three ports of the router. 
Here, the positive values of the measurements corresponds to the inflow to the router. 
The operation is divided into three phases: a transient from $t=1.0\,{\rm s}$ to $3.0\,{\rm s}$ including the mode switch of the router at $t=1.973\,{\rm s}$, and two steady states before and after the transient. 

\begin{figure}
    \centering
    \includegraphics[width=.9\linewidth]{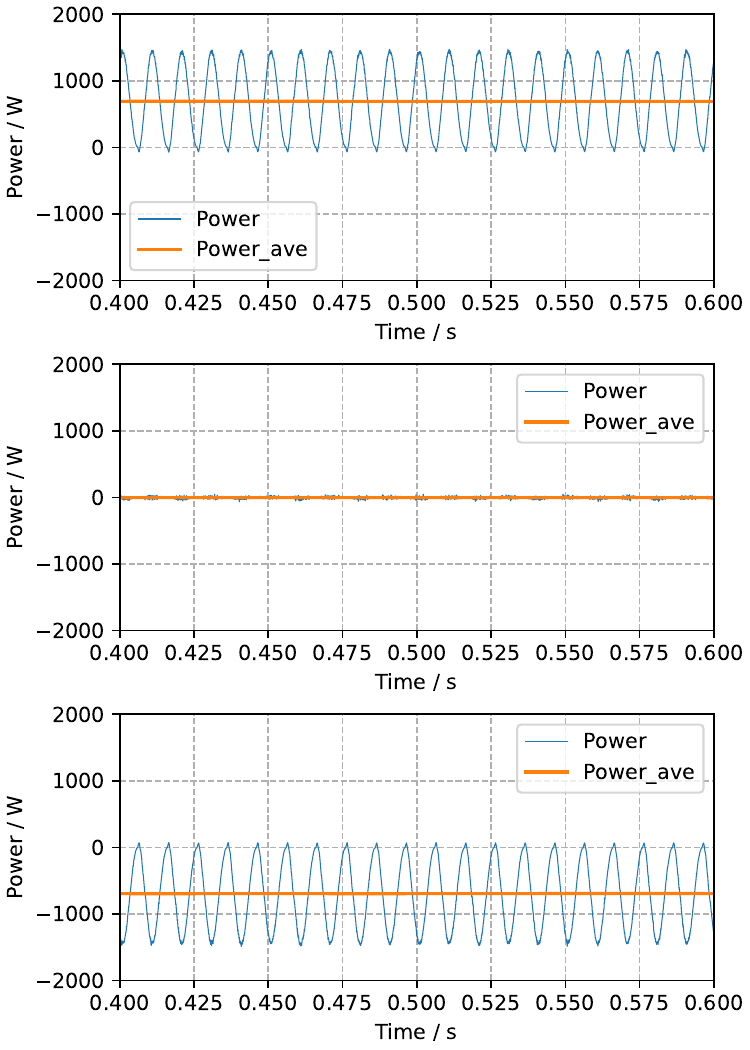}
    \caption{Enlarged view of \fref{fig:p123} during the first steady state. The meanings of top, middle, and bottom plots and colored lines therein are the same as \fref{fig:p123}.}
    \label{fig:p123_enlarged1}
\end{figure}

\begin{figure}
    \centering
    \includegraphics[width=.9\linewidth]{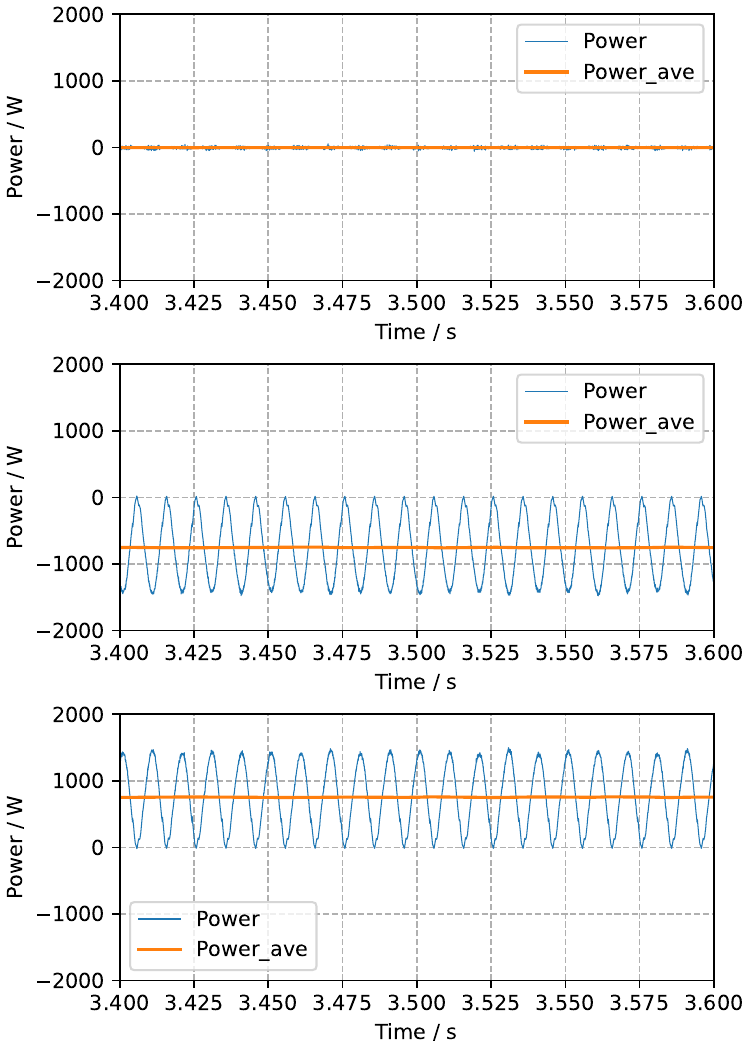}
    \caption{Enlarged view of \fref{fig:p123} during the second steady state. The meanings of top, middle, and bottom plots and colored lines therein are the same as \fref{fig:p123}.}
    \label{fig:p123_enlarged2}
\end{figure}

First, we see the results in the steady state operation. 
Before $t=1.0\,{\rm s}$, the power inflow is observed in port 1, the outflow in port 3, and no flow in port 2. 
This result is consistent with the scenario before the mode transition, namely that house 1 supplies surplus power to the battery and house 2 does not participate in the power sharing because it achieves the balance of supply and demand on its own. 
\fref{fig:p123_enlarged1} presents the enlarged view of the power waveforms in this steady state. 
The constant power supply from house 1 to battery is confirmed. 
Then, after $t=3.0\,{\rm s}$, the outflow is observed in port 2, inflow in port 3, and no flow in port 1.
This is in line with the designed operation of the router in mode 2, where house 2 is now short of power while house 1 achieves the balance of supply and demand. 
\fref{fig:p123_enlarged2} presents the enlarged view of the power waveforms in this steady state. 
The constant power supply from battery to house 2 is confirmed. 

\begin{figure}
    \centering
    \includegraphics[width=.9\linewidth]{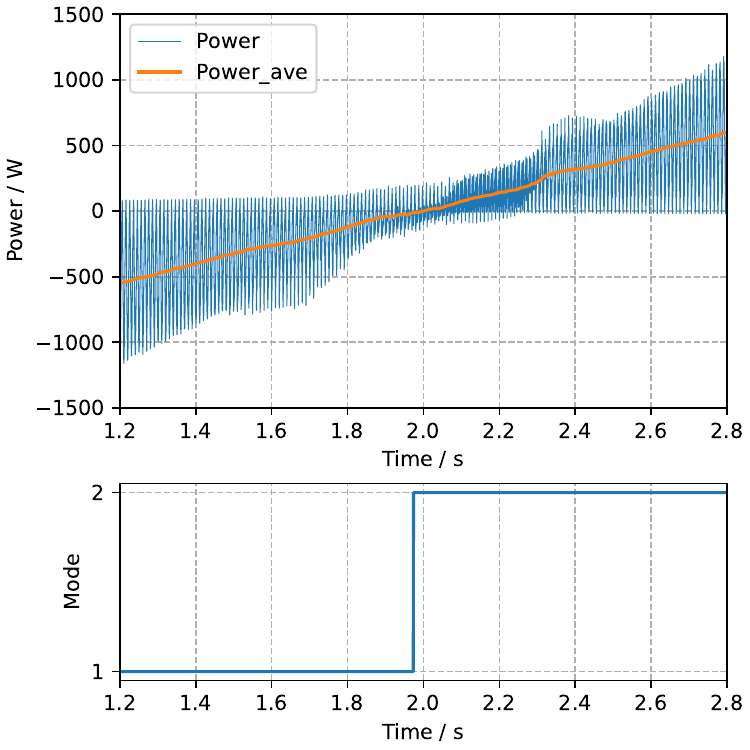}
    \caption{Power waveforms at port 3 (top) and router's mode transition (bottom). }
    \label{fig:p1_mode}
\end{figure}

Next, we move on to the results in the transient state. 
From $t=1.0\,{\rm s}$, the battery gradually reduces the amount of charging power to switch to the discharging operation. 
It should be noted that the output control of the battery is triggered by the external signal and the switching state of the router does not change at this time. 
Then, when the router detects the first zero-crossing of the power flow, the router alters the switching states from mode 1 to mode 2. 
\fref{fig:p1_mode} presents an enlarged view of the power flow to/from battery in port 3 around the time instance of the mode switch.  
After the switch, the battery smoothly transitions to the discharging operation and gradually increases the amount of output to reach the steady state.

\begin{figure}
    \centering
    \includegraphics[width=.9\linewidth]{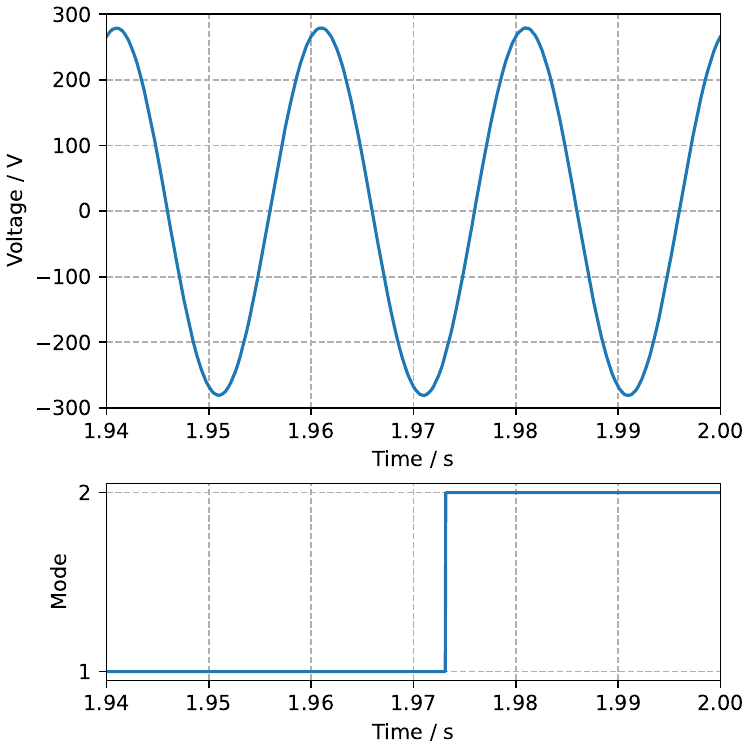}
    \caption{Bus voltage waveform and router's mode transition (bottom). }
    \label{fig:v1_mode}
\end{figure}

Lastly, we confirm the effect of the mode switching on the bus voltage. 
\fref{fig:v1_mode} presents the enlarged view of the bus voltage around the mode switching. 
The result clearly shows that no surge or distortion is induced by the circuit switching of the router. 

\section{Conclusions}
In this paper, we experimentally demonstrated the bidirectional power routing in a situation where the output polarity of the battery dynamically changed. 
The newly proposed algorithm to switch the router's connection based on the onboard power flow monitoring was proven to successfully detect the appropriate switching timing.
These results support the feasibility of the energy management system with bidirectional power router. 

\section*{Acknowledgment}

The authors thank Professor Takashi Hikihara of Kyoto University for his invaluable advice. 

\bibliographystyle{IEEEtran}
\bibliography{bidir-2025}

\end{document}